\documentclass[aps,prl,reprint,superscriptaddress,showpacs]{revtex4-1} 
\usepackage{graphicx}  
\usepackage{dcolumn}   
\usepackage{amssymb}
\usepackage{physics}
\usepackage{natbib}
\usepackage{color}

\usepackage[textsize=tiny]{todonotes}

\begin{document}

\title{Anharmonicity of a Gatemon Qubit with a Few-Mode Josephson Junction}

\author{A.~Kringh{\o}j}
\affiliation{Center for Quantum Devices and Station Q Copenhagen, Niels Bohr Institute, University of Copenhagen, Copenhagen, Denmark}
\author{L.~Casparis}
\affiliation{Center for Quantum Devices and Station Q Copenhagen, Niels Bohr Institute, University of Copenhagen, Copenhagen, Denmark}
\author{M.~Hell}
\affiliation{Center for Quantum Devices and Station Q Copenhagen, Niels Bohr Institute, University of Copenhagen, Copenhagen, Denmark}
\affiliation{Division of Solid State Physics and NanoLund,
Lund University, Lund, Sweden}
\author{T.~W.~Larsen}
\affiliation{Center for Quantum Devices and Station Q Copenhagen, Niels Bohr Institute, University of Copenhagen, Copenhagen, Denmark}
\author{F.~Kuemmeth}
\affiliation{Center for Quantum Devices and Station Q Copenhagen, Niels Bohr Institute, University of Copenhagen, Copenhagen, Denmark}
\author{M.~Leijnse}
\affiliation{Center for Quantum Devices and Station Q Copenhagen, Niels Bohr Institute, University of Copenhagen, Copenhagen, Denmark}
\affiliation{Division of Solid State Physics and NanoLund,
Lund University, Lund, Sweden}
\author{K.~Flensberg}
\affiliation{Center for Quantum Devices and Station Q Copenhagen, Niels Bohr Institute, University of Copenhagen, Copenhagen, Denmark}
\author{P.~Krogstrup}
\affiliation{Center for Quantum Devices and Station Q Copenhagen, Niels Bohr Institute, University of Copenhagen, Copenhagen, Denmark}
\author{J.~Nyg\r{a}rd}
\affiliation{Center for Quantum Devices and Station Q Copenhagen, Niels Bohr Institute, University of Copenhagen, Copenhagen, Denmark}
\affiliation{Nano-Science Center, Niels Bohr Institute, University of Copenhagen, Copenhagen, Denmark}
\author{K.~D.~Petersson}
\affiliation{Center for Quantum Devices and Station Q Copenhagen, Niels Bohr Institute, University of Copenhagen, Copenhagen, Denmark}
\author{C.~M.~Marcus}
\affiliation{Center for Quantum Devices and Station Q Copenhagen, Niels Bohr Institute, University of Copenhagen, Copenhagen, Denmark}

\begin{abstract}
Coherent operation of gate-voltage-controlled hybrid transmon qubits (gatemons) based on semiconductor nanowires was recently demonstrated. Here we experimentally investigate the anharmonicity in epitaxial InAs-Al Josephson junctions, a key parameter for their use as a qubit. Anharmonicity is found to be reduced by roughly a factor of two compared to conventional metallic junctions, and dependent on gate voltage. Experimental results are consistent with a theoretical model, indicating that Josephson coupling is mediated by a small number of highly transmitting modes in the semiconductor junction.  
\end{abstract}

\maketitle

The nonlinear inductance of the Josephson junction (JJ) makes the transmon qubit an anharmonic oscillator, allowing the lowest two energy levels to be selectively addressed \cite{koch_2007, houck_2007, devoret_2013}. The anharmonicity $\alpha = E_{12} - E_{01}$, where $E_{ij}$ is the energy difference between energy states {\em j} and {\em i}, is a critical qubit design parameter, determining, for instance, the minimum pulse duration $\sim \hbar/|\alpha|$ needed to avoid leakage into noncomputational states. Transmons have recently demonstrated one and two qubit gate fidelities exceeding 0.99 in multi-qubit devices \cite{barends_2014, kelly_2015, sheldon_2016}.

Almost without exception, transmons are based on superconductor-insulator-superconductor (SIS) junctions that use a thin insulating barrier (typically Al$_2$O$_3$) between metallic superconducting leads~\cite{paik_2011}. SIS junctions are well described by a non-harmonic (cosine) energy-phase relation, $V_{SIS} = -E_J \, \textrm{cos} (\phi)$, where $E_J$ is the Josephson coupling energy and $\phi$ is the  phase difference across the junction~\cite{girvin_2014}. The  inverse inductance correspondingly depends on phase, $L_{SIS}^{-1} = (2e/\hbar)^2 d^2V_{SIS}/d\phi^2 = (2e/\hbar)^2 E_J \textrm{cos} (\phi)$. Other types of JJs, with weak links separating superconducting electrodes made from narrow superconducting constrictions, normal metal, or a semiconductor~\cite{golubov_2004, bretheau_2013,doh_2005} have energy-phase relations that differ from the cosine form. Coherent operation of one- and two-qubit circuits using superconductor-semiconductor-superconductor (S-Sm-S) junctions---called gatemons due to their gate-voltage controlled $E_J$---was recently demonstrated using an InAs nanowire (NW) with epitaxial Al~\cite{larsen_2015,casparis_2016}. In those experiments, it was noted that $\alpha$ was roughly a factor of two smaller than what one would expect for an SIS junction with the same operating parameters, but the origin and parameter dependence of this discrepancy was not investigated. 

Other experiments have investigated an S-Sm-S JJ in a two-junction loop~\cite{delange_2015}. Near one-half flux quantum through the loop, the anharmonic spectrum revealed signatures of a noncosinusoidal energy-phase relation in the S-Sm-S junction. More recently, nonsinusoidal current-phase relations of nanowire S-Sm-S junctions were directly measured from the diamagnetic response of mesoscopic rings interrupted by single S-Sm-S junctions~\cite{spanton}. 

In this Letter, we investigate anharmonicity as well as departure from the standard (SIS) cosine energy-phase relation in a nanowire-based gatemon qubit. We observe that anharmonicity depends on gate voltage and is lower than the corresponding SIS junction with comparable $E_J$. By comparing anharmonicity data to a model of Josephson junctions with few conduction channels, we are able to determine the number of conducting channels contributing to the Josephson current, with values in the range 1--3 channels, depending on gate voltage. 

\begin{figure}[!b]\vspace{-4mm}
    \includegraphics[width=1\columnwidth]{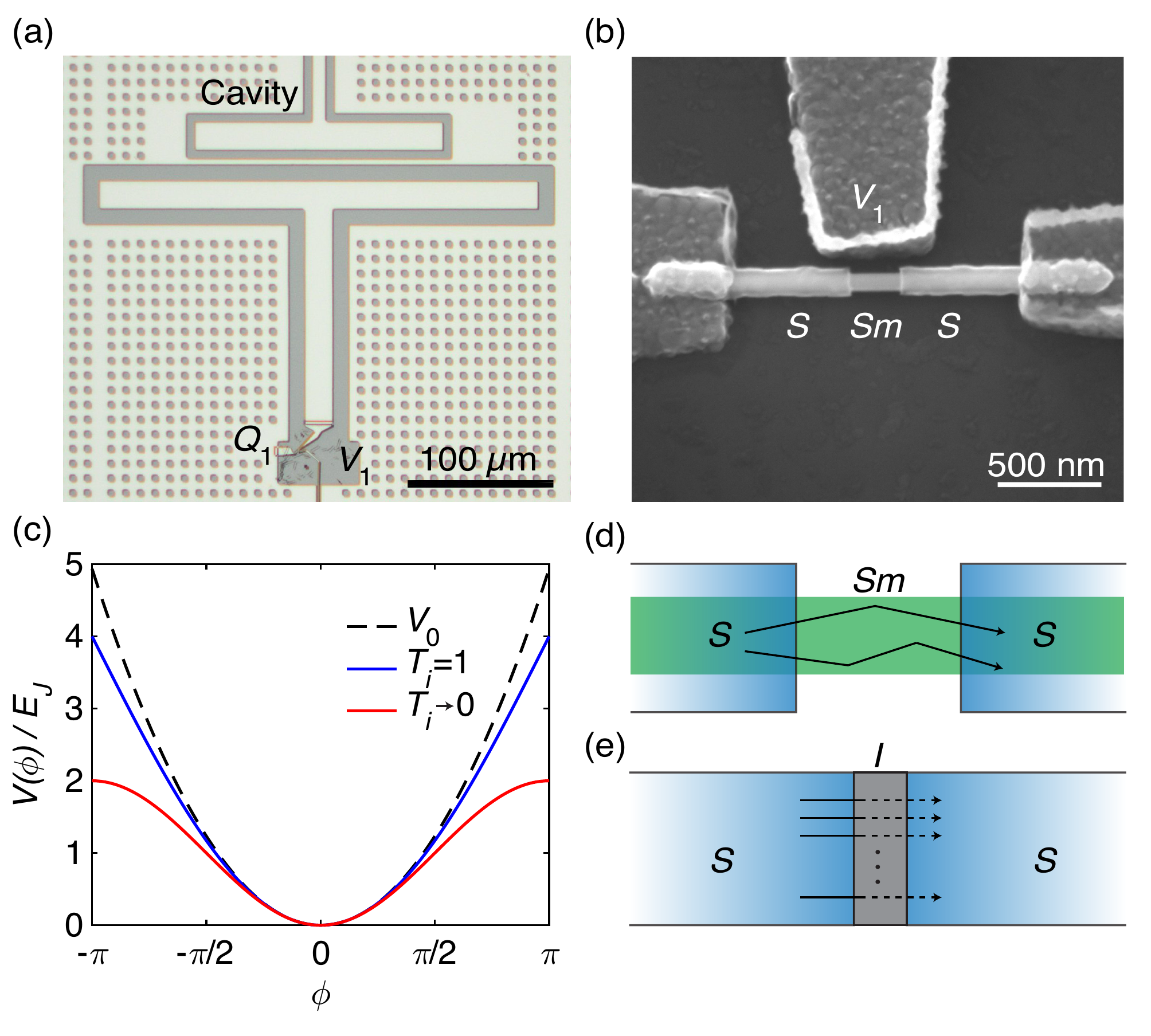}
    \caption{\vspace{-0mm} Qubit device and Josephson potential. (a) Optical micrograph of one of the qubits, Q1, in the two qubit device. Each qubit consists of a T-shaped island shunted to ground via an InAs/Al NW JJ. The two qubits are designed to be nominally identical and are both coupled to individual readout resonators. (b) Scanning electron micrograph of the S-Sm-S JJ for $Q_1$. The JJ features an InAs NW with high transparancy epitaxial Al contacts. The voltage, $V_1$, on the side gate modulates the density of carriers in the NW, allowing the Josephson potential to be modified. (c) The normalized Josephson potential $V(\phi)$ in the limits of $T_i=1$ (blue solid curve) and $T_i \to 0$ (red solid).  For comparison, a harmonic potential $V_0$ is also shown (black dashed). (d) Sketch illustrating a NW S-Sm-S JJ with a few highly transmitting channels in a quasiballistic regime as expected in the NW. (e) Sketch of the conventional SIS tunnel junction with many low transmitting channels.
    }
    \label{device}
\end{figure}

The gatemon qubit consists of a superconducting island with charging energy $E_C$, connected to ground via a single JJ made from a $L \sim 200$ nm segment of bare InAs NW, with superconducting leads proximitized by a full shell of epitaxial Al~\cite{krogstrup_2015} (details below). The mean free path in InAs NWs is typically $l \sim 100$ nm~\cite{chuang_2013, chang_2015}.  Taking the Al superconducting coherence length $\xi_0 \sim 1600$ nm~\cite{kittel_2005} gives a junction coherence length $\xi = \sqrt{\xi_0 l} \sim 400$ nm. In this regime, transport is weakly diffusive ($L > l$) but close to the so-called short-junction limit ($L \ll \xi$)~\cite{golubov_2004}. In the short-junction regime, originally considered by Beenakker for the case of a metal junction, multiple conduction channels are characterized by their transmission eigenvalues $\{T_i\}$ \cite{beenakker_1993}. Within this model, charge transport across the junction occurs via Andreev processes at each S-Sm interface. For each transmission channel, multiple Andreev reflections between the two interfaces result in a pair of discrete subgap states or Andreev bound states, each with ground state energy $-\Delta \sqrt{1-T_i \textrm{sin}^2(\phi/2)}$, where $\Delta$ is the induced superconducting gap in the leads~\cite{bretheau_2013, janvier_2016,woerkom_2016}. Summing over all conduction channels gives the Josephson potential
\begin{align*}
V(\hat{\phi})=-\Delta\sum_i \sqrt{1-T_i\sin^2(\hat{\phi}/2}),
\end{align*}
where $\hat{\phi}$ is the superconducting phase-difference operator.

The gatemon qubit is operated in the transmon regime, $E_J/E_C \gg 1$, where sensitivity to offset charge of the island is exponentially suppressed~\cite{koch_2007}. Omitting the offset charge, the effective Hamiltonian is given by
\begin{align*}
\hat{H}=4E_C\hat{n}^2 + V(\hat{\phi}),
\end{align*}
where $\hat{n}$ is the island Cooper pair number operator, conjugate to $\hat{\phi}$. The qubit transition frequency is given by the Josephson plasma frequency, $f_{01}\approx\sqrt{8E_CE_J}/h$.

To examine how anharmonicity, $\alpha$, depends on the channel transmission probabilities, we derive an expression for $\alpha$ by expanding $V(\hat{\phi})$ to $4th$ order in $\hat{\phi}$,
\begin{align*}
V(\hat{\phi}) & \approx \frac{\Delta}{4}\sum_i\left(\frac{ T_i}{2}\hat{\phi}^2-\frac{T_i}{24}(1 - \frac{3}{4}T_i)\hat{\phi}^4\right)\\
& = E_J\frac{\hat{\phi}^2}{2} - E_J\left(1-\frac{3\sum T_i^2}{4\sum T_i}\right)\frac{\hat{\phi}^4}{24},
\end{align*}
where the constant term has been omitted and $E_J=\frac{\Delta}{4}\sum T_i$~\cite{koch_2007,girvin_2014, reed_2013}. Here, the $\hat{\phi}^2$-term has the same form as the potential $V_0(\hat{\phi})=E_J\frac{\hat{\phi}^2}{2}$ in the harmonic oscillator Hamiltonian $\hat{H_0}=4E_C\hat{n}^2+V_0(\hat{\phi})$. Treating $V'(\hat{\phi})= - E_J\left(1-\frac{3\sum T_i^2}{4\sum T_i}\right)\frac{\hat{\phi}^4}{24}$ as a perturbation to $\hat{H_0}$ allows us to calculate the corrections to the harmonic transition energies. Evaluating the perturbation matrix elements $\bra{i}V'(\hat{\phi})\ket{i}$ for $i=0,1,2$ leads to
\begin{align*}
\alpha\approx-E_C\left(1-\frac{3\sum T_i^2}{4\sum T_i}\right).
\end{align*}
In the limit of $T_i\rightarrow0$ for all $i$, $\alpha=-E_C$ as is the case for transmons with SIS JJs~\cite{koch_2007}. For $T_i = 1$, $\alpha = -E_C/4$, giving a reduced qubit nonlinearity compared to the SIS JJ case.

Figure \ref{device}(c) illustrates the connection between channel transmissions and anharmonicity by comparing the Josephson potential in two limiting cases, $T_i = 1$ and $T_i \to 0$, to a harmonic potential ($\alpha = 0$). The case $T_i \to 0$ yields a $-\textrm{cos}(\phi)$ potential, corresponding to an SIS tunnel barrier with many low-transmission channels [Fig.~\ref{device}(e)]. The ballistic case, $T_i = 1$, yields a $-\textrm{cos}(\phi/2)$ potential, which more closely resembles a harmonic potential. For NW S-Sm-S JJs with quasiballistic transport dominated by a few channels [Fig.~\ref{device}(d)], one expects and observes behavior between these two limits. 

\begin{figure}
    \centering
        \includegraphics[width=1\columnwidth]{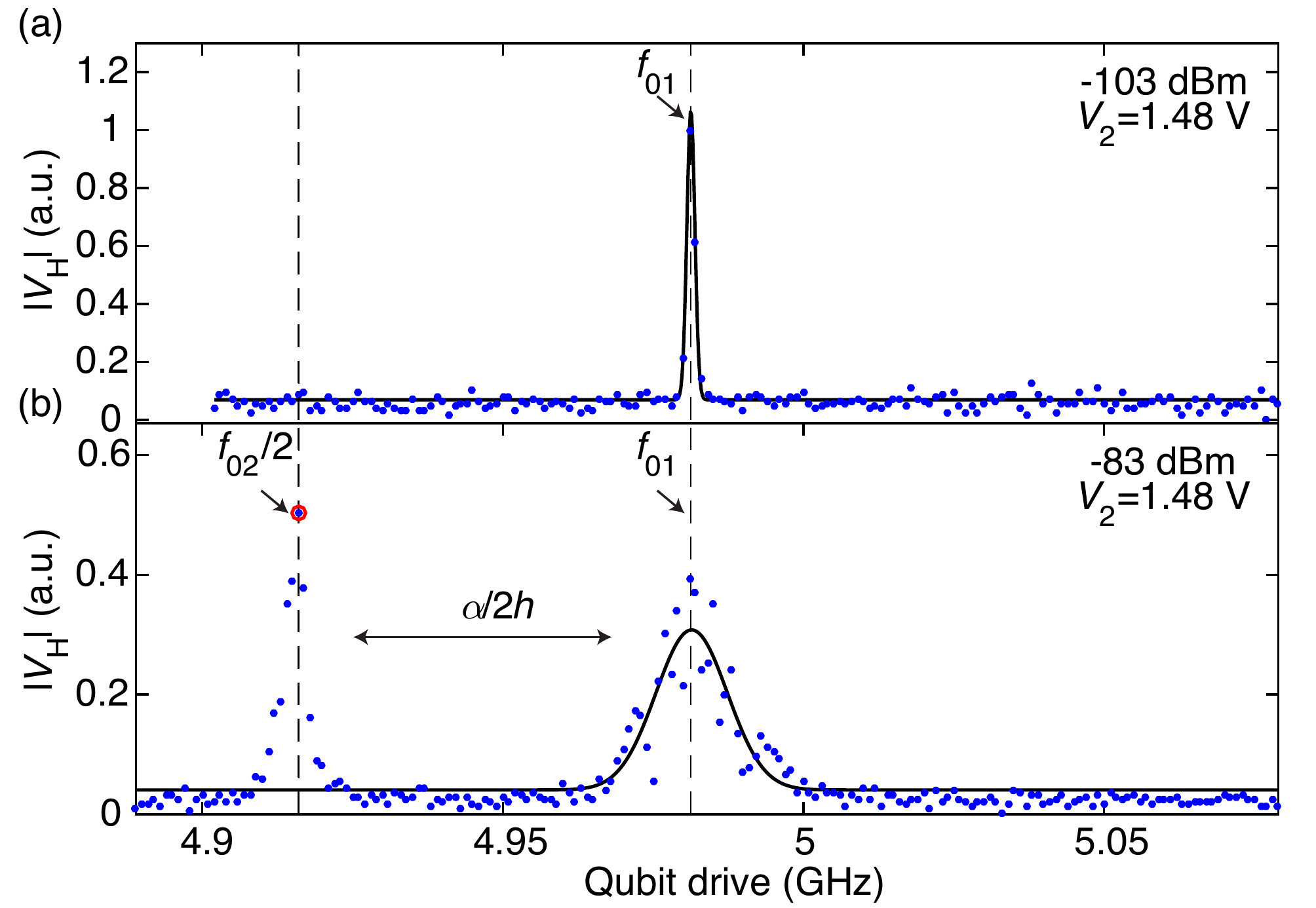}\vspace{-2mm}
    \caption{\vspace{-0mm} Spectroscopy scans to probe the anharmonicity. (a) The qubit is driven by a $-103$ dBm microwave pulse, which excites the qubit at the transition frequency $f_{01}$. By measuring the qubit-state-dependent demodulated cavity response $|V_\text{H}|$, $f_{01}$ can be determined. The data points (blue) are fitted to a Gaussian (solid black curve) to locate $f_{01}$. (b) After identifying $f_{01}$, the two photon $\ket{0}\rightarrow\ket{2}$ transition frequency $f_{02}/2$ is probed in a spectroscopy scan at $-83$ dBm. $f_{01}$ is extracted by fitting a Gaussian around the value found in (a) and $f_{02}/2$ is identified as the maximum value of the second peak as labelled in red. $\alpha/2h=f_{02}/2-f_{01}$ is indicated with the horizontal arrow.}
    \label{spectroscopy example}
\end{figure}

Experiments were carried out using a two-qubit device, similar to device in Ref.~\cite{casparis_2016}. Figures \ref{device}(a) and (b) show one of the qubits and its NW JJ. Control lines and qubit islands are lithographically defined on a 100 nm thick Al film evaporated on a high resistivity Si substrate. The JJ is constructed from a NW with a $\sim$75 nm diameter InAs core and a $\sim$30 nm thick epitaxial Al shell~\cite{krogstrup_2015}, where a $\sim$200 nm segment of the shell is removed by wet etching~\cite{larsen_2015, casparis_2016}. The two qubits, denoted Q1 and Q2, are coupled with strength $g/2\pi\sim80$ MHz to individual superconducting $\lambda/4$ resonators with resonance frequencies $f_{\text{C1}}\approx7.66$ GHz, $f_{\text{C2}}\approx7.72$ GHz. Multiplexed dispersive readout is performed through a common transmission line~\cite{barends_2013}, using a superconducting travelling wave parametric amplifier to improve the signal-to-noise and reduce the acquisition time~\cite{macklin_2015}. Coherence measurements show qubit lifetimes and inhomogeneous dephasing times, $T_1,T_2^* \sim$ 1--2 $\mu$s. Both quasi-two-dimensional and fully three-dimensional electrostatic simulations~\cite{comsol, ansys} yield $E_C/h=240$ MHz, taking Si permittivity $\epsilon=11.7$. 

Anharmonicity is measured by first locating the qubit transition frequency $f_{01}$ in a low-power scan (typically $\sim$\,$-100$ dBm at the sample). This is done by applying a microwave excitation with a pulse length of 1 $\mu$s through a control line capacitively coupled to the qubit island. The microwave pulse excites the qubit into a mixed state when applied at $f_{01}$, directly detectable in the demodulated cavity response $|V_\text{H}|$, as shown in Fig.~\ref{spectroscopy example}(a). Repeating the scan at higher power ($\sim$\,$-80$ dBm) allows both $f_{01}$ and the two-photon $\ket{0}\rightarrow\ket{2}$ transition frequency $f_{02}/2$ to be measured simultaneously, as shown in Fig.~\ref{spectroscopy example}(b). Frequency $f_{01}$ is extracted from a Gaussian fit to the $\ket{0}\rightarrow\ket{1}$ transition peak, while $f_{02}/2$ is taken to be the maximum value of the $\ket{0}\rightarrow\ket{2}$ peak. Anharmonicity is then given by $\alpha=2h\left(f_{02}/2-f_{01}\right)$.

\begin{figure}
    \centering
        \includegraphics[width=1\columnwidth]{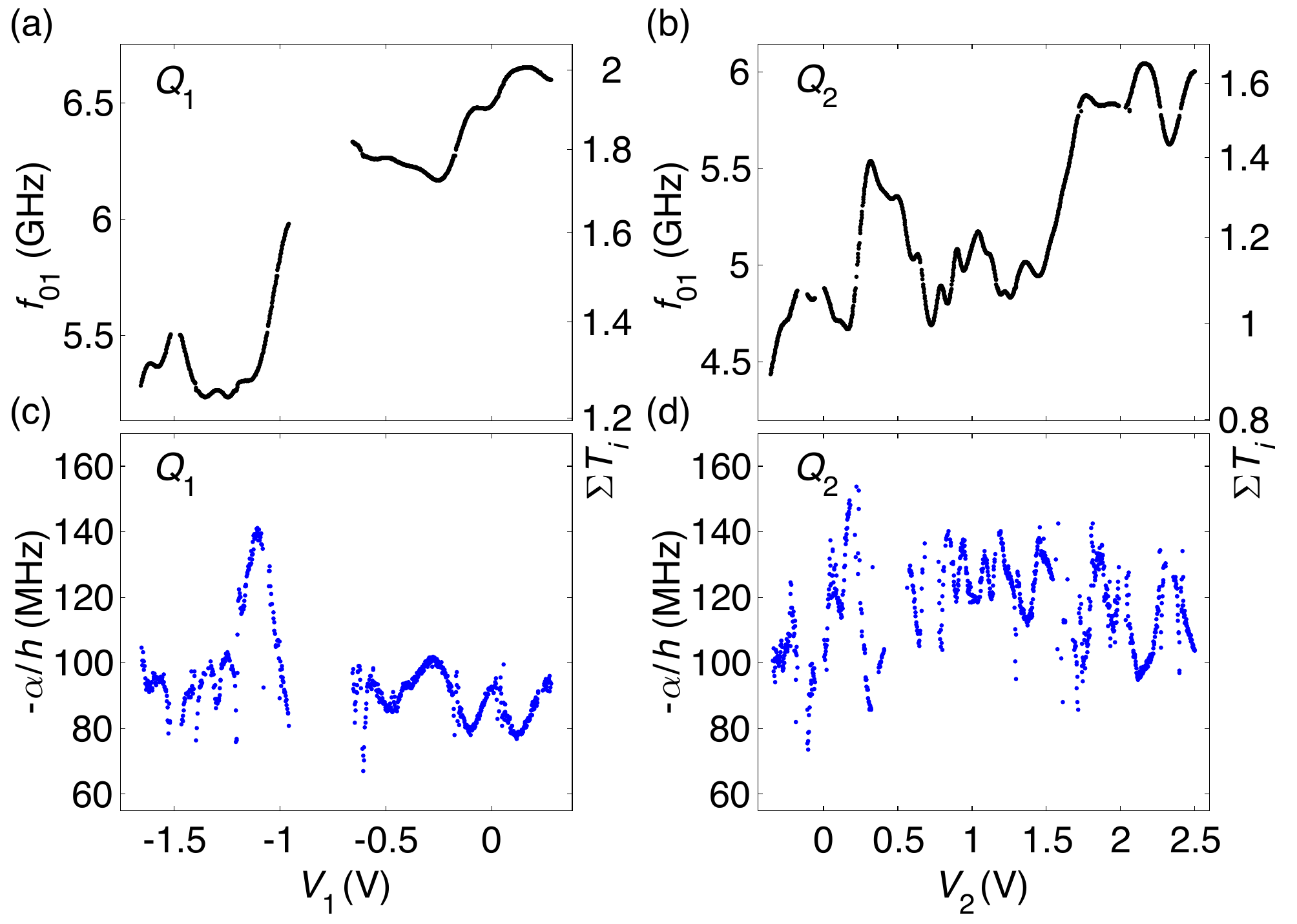}\vspace{-4mm}
    \caption{\vspace{-0mm} Results of the spectroscopy and anharmonicity measurements. (a) [(b)] Results of spectroscopy measurements of $f_{01}$ for varying gate voltage $V_{1}$ ($V_{2}$) on $Q_{1}$ ($Q_{2}$). The right axis indicates the total transmission $\sum T_i$ as converted from $f_{01}$ (see text). (c) [(d)] Results for $\alpha/h$ measured for $Q_{1}$ ($Q_{2}$) as a function of gate voltage, $V_{1}$ ($V_{2}$).  
    }
    \label{Results}
\end{figure}

Tuneability of the junction allows $f_{01}$ and $\alpha$ to be measured for different sets of channel transmissions, $\{T_i\}$, by performing  spectroscopy at different gate voltages, as shown in Fig.~\ref{Results}.  The right axes in Figs.~\ref{Results} (a,b) show $\sum T_i = (hf_{01})^2/2\Delta E_C$, taking $E_C$ from electrostatic modeling and $\Delta=190$ $\mu$eV, from measurements of similar NWs \cite{chang_2015}. Nonmonotonic gate dependence presumably reflects standing waves in the junction due to disorder, as discussed previously~\cite{larsen_2015, casparis_2016}. Figure \ref{Results}(c,d) shows anharmonicity $\alpha$ as a function of gate voltages. Both qubits show reduced anharmonicity compared to the corresponding for SIS value, $|\alpha| = E_C= 240$ MHz$\times h$, with sizable fluctuations with gate voltage. 

\begin{figure}
    \centering
        \hspace{-2mm}\includegraphics[width=1\columnwidth]{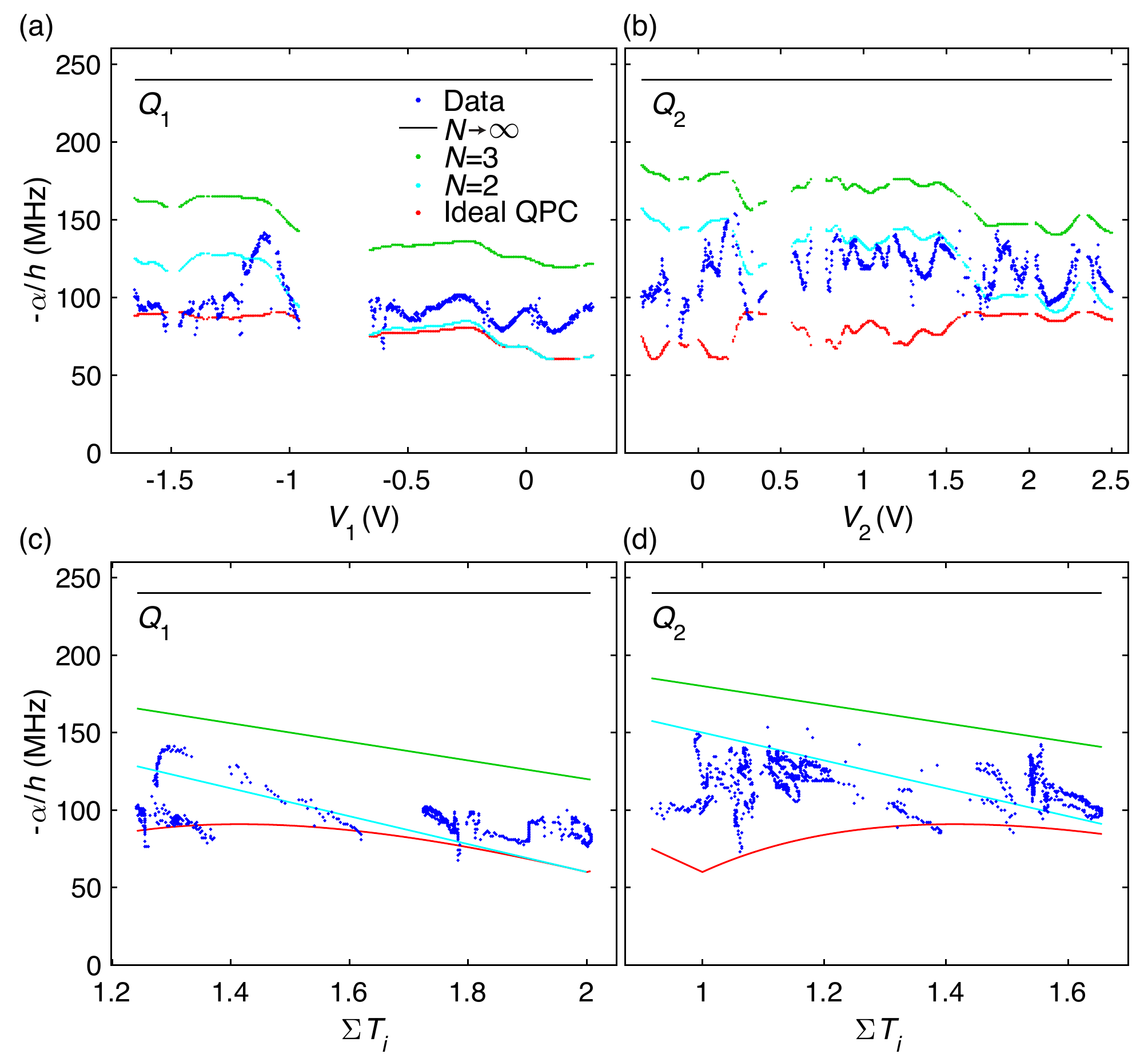}\vspace{-4mm}
    \caption{Comparison of the anharmonicity data (dark blue) to our model with four different channel transmission distributions for the JJ. Three of the distributions assume $N$ equally distributed channels plotted for $N=2$ (light blue), $N=3$ (green) and $N \to \infty$ (black). The fourth model data set (red) is for an ``Ideal QPC" distribution (see main text for further details).  (a) [(b)] $\alpha$ as a function of $V_1$ ($V_2$) compared with the different models. (c) [(d)] $\alpha$ plotted parametrically against $\sum T_i$ for $Q_1$ ($Q_2$), as determined from Fig.~\ref{Results}(a)[(b)].
    }
    \label{Theory curves}\vspace{-4mm}
\end{figure}

Spectroscopy data along with model calculations for several different distributions for $\{T_i\}$ are shown in Fig.~\ref{Theory curves}, as functions of both gate voltage and total transmission, $\sum T_i$, extracted from Figs.~\ref{Results}(a) and (b). Theoretical plots show the model for three cases of equal transmission probability, $T$, in each channel, $\alpha=-E_C\left(1-\frac{3}{4}T\right)=-E_C\left(1-\frac{3E_J}{\Delta N}\right)$ for different number of participating channels, $N=2$, 3, and $ \infty$. A fourth model (``Ideal QPC'') assumes that the $\{T_i\}$ are maximally packed for a given total transmission, that is, channels are filled in a staircase with at most one  partially transmitting channel, setting a lower bound on anharmonicity. Comparing experimental data for both qubits to these four cases indicates that  transmission involves between one and three channels, for all measured gate voltages. 

Measured values of anharmonicity for the gatemon are reduced by a factor of $\sim$2 compared to corresponding transmons with SIS junctions. As a consequence, control pulses must be a factor of $\sim$2 slower for the gatemon to avoid state leakage. SIS-based transmons are typically designed with $E_C/h = 200-300$ MHz to allow for fast control pulses, in the few-ns regime, while maintaining $E_J/E_C \gg 1$ to ensure dephasing due to charge noise and quasiparticle poisoning is suppressed~\cite{schreier_2008}. This regime may not be optimal for the gatemon, however, and it may be possible to increase $E_C$ to allow faster control while remaining insensitive to charge fluctuations in the island. This is because when any channel transmission approaches unity, energy dispersion with charge is predicted to vanish~\cite{averin_1999}. Similarly, recent experiments with a normal metal island have shown the quenching of charging quantization in the limit of a ballistic channel~\cite{jezouin_2016}. In future work we will exploit this reduced (and in principle vanishing) dispersion to find optimal $E_J/E_C$ ratio for gatemons. 

In summary, we have measured anharmonicity of a gatemon qubit, yielding information about the set of transmissions of the few participating channels in the semiconductor junction. Our results indicate that three or fewer channels significantly participate in transport, depending on gate voltage. We note that one may further exploit the noncosine form of the energy-phase relations to create novel superconducting elements, such as $\textrm{cos} (2\phi)$ junctions, which would enable new types of qubits that are intrinsically protected against sources of decoherence~\cite{doucot_2012, bell_2014, delange_2015}. 

\begin{acknowledgments}
This work was supported by Microsoft, the U.S. Army Research Office, the Danish National Research Foundation, the Crafoord Foundation (M.H.~and M.L.), the Swedish Research Council (M.L.), and the Villum Foundation (C.M.M.). The traveling wave parametric amplifier used in this experiment was provided by MIT Lincoln Laboratory and Irfan Siddiqi Quantum Consulting (ISQC), LLC via sponsorship from the US Government.
\end{acknowledgments}


%

\end{document}